\documentclass[10pt]{article}
\usepackage{graphicx}
\usepackage{latexsym}
\usepackage{amssymb}
\usepackage{amsfonts}
\usepackage{amsmath}
\usepackage{theorem}
\newtheorem{theorem}{Theorem}
\newtheorem{lemma}{Lemma}
\newtheorem{corollary}{Corollary}
\newtheorem{definition}{Definition}

\newtheorem{proposition}{Proposition}
\makeatletter
\newcommand{\contraction}[5][1ex]{%
  \mathchoice
    {\contraction@\displaystyle{#2}{#3}{#4}{#5}{#1}}%
    {\contraction@\textstyle{#2}{#3}{#4}{#5}{#1}}%
    {\contraction@\scriptstyle{#2}{#3}{#4}{#5}{#1}}%
    {\contraction@\scriptscriptstyle{#2}{#3}{#4}{#5}{#1}}}%
\newcommand{\contraction@}[6]{%
  \setbox0=\hbox{$#1#2$}%
  \setbox2=\hbox{$#1#3$}%
  \setbox4=\hbox{$#1#4$}%
  \setbox6=\hbox{$#1#5$}%
  \dimen0=\wd2%
  \advance\dimen0 by \wd6%
  \divide\dimen0 by 2%
  \advance\dimen0 by \wd4%
  \vbox{%
    \hbox to 0pt{%
      \kern \wd0%
      \kern 0.5\wd2%
      \contraction@@{\dimen0}{#6}%
      \hss}%
    \vskip 0.5ex
    \vskip\ht2}}

\newcommand{\contraction@@}[3][0.05em]{%
  \hbox{%
    \vrule width #1 height 0pt depth #3%
    \vrule width #2 height 0pt depth #1%
    \vrule width #1 height 0pt depth #3%
    \relax}}
\makeatother

\begin{document}

\title{\bf Positive-Overlap Transition and Critical Exponents in Mean Field
Spin Glasses}
\author{Alessandra Agostini \footnote{School of Mathematics, Cardiff
University, Senghennydd Road, Cardiff, CF24 4AG, Wales, United
Kingdom {\tt<agostinia@cf.ac.uk>}}, Adriano Barra\footnote{King's
College London, Department of Mathematics, Strand, London WC2R 2LS,
United Kingdom, and Dipartimento di Fisica, Universit\`a di Roma
``La Sapienza'' Piazzale Aldo Moro 2, 00185 Roma, Italy,
{\tt<Adriano.Barra@roma1.infn.it>}}, Luca De Sanctis \footnote{ICTP,
Strada Costiera 11, 34014 Trieste, Italy, {\tt<lde\_sanc@ictp.it>}}
 }

\def\be{\begin{equation}}
\def\ee{\end{equation}}
\def\bes{\begin{equation*}}
\def\ees{\end{equation*}}
\def\bc{\begin{center}}
\def\ec{\end{center}}

\maketitle

\begin{abstract}
In this paper we obtain two results for the
Sherrington-Kirkpatrick (SK) model, and we show that they
both emerge from a single approach. First, we prove
that the average of the
overlap takes positive values when it
is non zero. More specificly, the average of the overlap,
which is naively expected to take values
in the whole interval $[-1,+1]$, becomes positive if we ``first''
apply an external field, so to destroy the gauge invariance of the
model, and ``then'' remove it in the thermodynamic limit. 
This phenomenon emerges at the critical point.
This first result is weaker than the one obtained by Talagrand 
(not limited to the average of the overlap), but we show here
that, at least in average, the overlap is proven to be non-negative 
with no use of the Ghirlanda-Guerra identities. The latter are instead
needed to obtain the second result, which is the
control the behavior of the overlap at the critical point: 
we find the critical exponents of all the overlap correlation
functions. 
\end{abstract}

\section{Introduction}

Mean field spin glass models are a very hard ``challenge for
mathematicians'' \cite{talabook}. In fact, despite the important
fact that the free energy is known (predicted by Parisi and
rigorously proven by Talagrand), very little is proven
about their organization into pure states at low temperature
except for high temperature regimes.
Only a sequence of brilliant intuitions of Parisi (see \cite{mpv}
and references therein) provided a proposal for a theory of
disordered systems, which is comforted by many numerical studies
(see for example \cite{enzo}), but many of the intuitions still
lack a rigorous proof. Among the many stimulating
suggestions, Parisi prescribed the positivity of the overlap
below the critical temperature, which
breaks the symmetry by which the overlap should be trivially zero
(in absence of an external field). More
precisely, the average of the overlap is zero when there is no
external field, by spin-flip symmetry. If an external field is
applied, the spin-flip symmetry is broken and the average of the
overlap is non-zero. When the external field is removed the average
of the overlap remains strictly positive, below the critical
temperature. This is what we prove here
(slightly more, to be precise, as we will use a random
external field as opposed to a constant one) in agreement with what
Parisi predicted. This result is weaker than what Talagrand proved
\cite{talabook}, namely that the overlap is non-negative
with the possible exception of a zero Lebesgue measure set of 
the parameters. This is shown to be a consequence of the 
Ghirlanda-Guerra identities \cite{gg}. Here we show 
that at least the weaker result can be proven in a 
``very physical'' and simple way. Moreover, the 
technique offers another result, of crucial physical relevance,
i.e. the critical exponents of the behavior of the overlaps at the
critical point.

We will focus on the following points
\begin{enumerate}
\item spin-flip symmetry breaking due to an external field;
\item strict positivity of the average of the overlap below the critical temperature;
\item gauge symmetry breaking due to an external field;
\item critical exponents in connection with the gauge symmetry breaking.
\end{enumerate}

The procedure we employed to prove these results
makes use of an iterative approach to expand the average of the
overlap in a power series of the strength of the external field.
Within a framework typical of cavity approaches, we also take
advantage of some known symmetries of the model, which we report 
together with the definition of the model in
section \ref{def}. 
The main results are presented in 
sections \ref{main} and \ref{exp}, where the positivity
of the overlap and the critical exponents are studied respectively.
Conclusive remarks are left for section \ref{final}.

\section{Model, notations, previous results}\label{def}

The Sherrington-Kirkpatrick (SK) model describes a system of $N$ binary
spins $\sigma_i , i \in (1 ,\ldots , N)$.
A configuration $\sigma$ of the system is then a map
\bes
\sigma :\{1,2,...,N\} \ni i \rightarrow \sigma_i \in
\{-1,+1\}\ .
\ees
The Hamiltonian of the model is defined to assign the following energy
to a given configuration $\sigma$:
\bes H_N(\sigma ; J)=
\frac{-1}{\sqrt{N}}\sum_{(i,j)}J_{ij}\sigma_i\sigma_j\ ,
\ees
where the sum ranges over all the $N(N-1)/2$ different couples of indices
$(i,j)$, and $J_{ij}$ are independ centered unit Gaussian random
variables.
The partition function is defined as usual by
\bes
Z_N(\beta ; J)=\sum_{\sigma}\exp(-\beta H(\sigma ; J))\ ,
\ees
while the Boltzmann-Gibbs expectation of an observable
$A:\sigma\to\mathbb{R}$ is
\be\label{gibbs}
\omega(A)=
\frac{1}{Z_N(\beta ; J)}\sum_{\sigma}A(\sigma)\exp(-\beta
H(\beta ; J))\ .
\ee
The global average (over the spins first and the noise in the
couplings afterwards) is
$\langle \cdot \rangle = \mathbb{E}\omega(\cdot)$,
where $\mathbb{E}$ denotes the expectation with respect
to all the (quenched) Gaussian variables.
The notation for the averages does not specify any dependence on the size
of the system, and we will occasionally use the same 
symbols in the thermodynamic limit too, specifying this when needed.
By $\Omega$ we denote the product measure ({\em replica measure})
of the needed number
of copies of $\omega$, which we will use when dealing
with functions of several configurations (replicas).
Notice that while $\Omega$ is factorized by definition,
$\langle\cdot\rangle$ is not, as we are just replicating
configurations, keeping for each the same disorder (i.e. Gaussian
couplings). Taking the same disorder results in coupling the various
replicas and therefore $\mathbb{E}$ destroys the factorization.
Given two replicas $\sigma^{(1)}$ and $\sigma^{(2)}$ we define the
overlap between them as
$$
q_{12}=\frac1N\sum_{i=1}^N \sigma_i^{(1)}\sigma_i^{(2)}\ ,
$$
sometimes simply written as $q$.
The pressure $\alpha_N(\beta)$ and the free energy per spin
$f_N(\beta)$ are defined as
\bes
\alpha_N(\beta)=-\beta f_N(\beta)=
\frac1N\mathbb{E}\ln Z_N(\beta)\ .
\ees
By a trivial (spin-flip) symmetry argument, one readily sees that (for 
all inverse temperature $\beta$)
$$
q_{12}\in[-1,1]\ ,\ \ \langle q_{12}\rangle=0\ .
$$
When an external field is applied, i.e. a term of the form
$\sqrt{x}\sum_{i=1}^N J_{i} \sigma_i$ for $J\in\mathcal{N}(0,1)$ and
$x\in\mathbb{R}_{+}$ is added to the Hamiltonian, the spin-flip
symmetry is broken and the overlap will be different from zero in
average. Parisi theory \cite{mpv} prescribes a phase transition such
that even an infinitesimal external field will force the overlap to
be supported on $[0,1]$, and the limit as $x \to 0^{+}$ of the
average of the overlap is positive:
$$
\lim_{x\to 0^{+}}\langle q_{12}\rangle>0\ , \ \beta > \beta_{c}
$$
where $\beta_{c}$ is the critical inverse temperature and 
clearly the dependence on $x$ of the measure
$\langle\cdot\rangle=\mathbb{E}\omega(\cdot)$
is here in the Boltzmannfaktor
in $\omega$, which includes the external field term
in addition to the original Hamiltonian.
For the sake of clarity, we will sometimes attach a sub $x$ to the
measures, when they include the external field.

\subsection{Some useful results}

The next two propositions, which we do not provide the easy proof
of, will be of precious help.
\begin{proposition}
The average $ \langle\cdot\rangle$ is invariant under replica
symmetry: let $g$ be any element of the group
$\mathbf{G}$ of permutations of $s$ objects, then
$$
\langle \phi(q_{ab}) \rangle =\langle \phi(q_{g(a)g(b)})\rangle
$$
for any function $\phi$ of two generic replicas and all $a,b\in\{1,\ldots ,s\}$.
\end{proposition}

\begin{proposition} The average $ \langle \cdot \rangle$
is invariant under gauge symmetry:
$$
\sigma_i^a \rightarrow \epsilon_a \sigma_i^a\ ,\
q_{ab}\rightarrow  \epsilon_a \epsilon_b q_{ab}
$$
being  $\epsilon_a,\epsilon_{b}=\pm1$ .
\end{proposition}
This symmetry is a consequence of the parity of the state $\omega$
and of the dichotomic nature of the spin, so it is obviously guaranteed by
the spin-flip symmetry of the Hamiltonian for each single replica
in $\Omega$.

We will have to compute several averages of polynomials of overlaps,
but will see that the average of certain polynomials are easy to compute.
That is why the following classification comes in handy.
\begin{definition} A polynomial function of some overlaps is said
\begin{itemize}
\item {\bf filled} if every replica appears an even number of times in it;
\item {\bf fillable} if it can be made filled multiplying it by exactly one overlap
of two replicas;
\item {\bf unfillable} if it is neither filled nor fillable.
\end{itemize}
\end{definition}
A consequence of the gauge invariance of the model is
that the average of non filled polynomials (fillable as well as
unfillable) is identically zero. The reason is that the absence of an
external field guarantees gauge invariance, and non-filled polynomials
are asymmetric in one of the overlaps and averages
of products of spins belonging to different replicas are zero by the
symmetries above.
Consequently, in this case the probability distribution of functions of
overlaps is uniquely determined by its restriction to the positive
semi-domain $[0,1]$.
The general case, at zero
external field, in which $q \in [-1,+1]$, is obtainable simply by
employing the gauge.

\section{Discontinuity of the order parameter}\label{main}

In order to establish the presence of a discontinuity
in the overlap due to an infinitesimal external field,
we employ some kind
of cavity method and stochastic stability techniques.
Namely, we add a random perturbation (as the one
described above) which can be equivalent
either to a deterministic external field or to the addition of
a new spin (depending on the choice of the parameters),
and we then study the
consequences of the spontaneous breaking of the spin-flip
and gauge symmetries.

\subsection{The external field and non-negativity of the average
of the overlap}

Before proceeding, in the next subsection, with the introduction
of the main techniques,
let us introduce the needed external field which breaks the spin-flip
symmetry of the Hamiltonian and hence of $\omega$.
Consider a family
$\{J_{i}\}$ of $N$ centered unit Gaussian independent random variables.
The expectation with respect to these (quenched)
variables will be included in $\mathbb{E}$. Let us introduce
consequently the following Boltzmannfaktor
$$
B(x) = \exp\bigg(-\beta H_{N}(\sigma; J)
+ \sqrt{x}\sum_{i=1}^{N}J_i \sigma_i\bigg)\ .
$$
We will here denote the corresponding Gibbs measure by $\omega_{x}$,
and $\langle \cdot \rangle_{x}\equiv \mathbb{E}\omega_{x}(\cdot)$.
In the following we will be interested in letting $x$ go to zero as
$N^{-1}$ while $N$ goes to infinity. This is the only way to recover
the correct normalization for a Hamiltonian of $N+1$ spins (that is
the proper weight of the Boltzmann factor $B(x)$ if the gauge
symmetry holds), and it is what is technically needed
to get our main result, as it will emerge.

Let us now report a result that generalizes slightly the analogous
one proven in \cite{barra}.
\begin{proposition}\label{xxx} In the $N \to \infty$ limit the
average $\langle \cdot \rangle_{x}$ of filled polynomials is
$x$-independent, i.e. $\langle
q_{12}q_{23}q_{13}\rangle_x = \langle q_{12}q_{23}q_{13}\rangle$,
when integrated over any interval $[\beta_{1},\beta_{2}]\ni \beta$,
i.e. with the exclusion of a zero Lebesgue measure set of
values of $\beta$.
\end{proposition}
We are not going to prove this proposition, as it obtainable as a special
case of the next one, which will be proven here.
\begin{theorem}\label{qqq}
Let $Q_{ab}$ be a fillable polynomial of some overlaps
(this means that  $ q_{ab}Q_{ab}$ is filled). Then
$$
\lim_{N \rightarrow \infty}\lim_{x\rightarrow\beta^2/N}\langle
Q_{ab}\rangle_{x}= \langle q_{ab}Q_{ab} \rangle\ ,
$$
where the right hand side too is understood to be taken
in the thermodynamic limit.
\end{theorem}
The proof requires the following
\begin{lemma}\label{n+1} 
Let  $\omega(.)$ and $\omega_x(.)$ be the
states defined respectively by the canonical partition function and
by the extended one; if we consider the ensemble of index
$\{i_1,..,i_r\}$ with $r\in [1,N]$, then in the limit $x=\beta^2/N$
the following relation holds:
$$
\lim_{x \rightarrow
\beta^2/N}\omega_{N,x}(\sigma_{i_1},...,\sigma_{i_r})=
\omega_{N+1}(\sigma_{i_1},...,\sigma_{i_r},\sigma_{N+1}^r)
+O(\frac{1}{N} )\ .
$$
\end{lemma}
In this statement $r$ is an exponent, not a replica index, so if $r$ is
even $\sigma_{N+1}^r=1$, while if it is odd
$\sigma_{N+1}^r=\sigma_{N+1}$ and we emphasized 
the dependence on $N$ or $N+1$ in the
two averages for clarity.\\
\textbf{Proof}
For the sake of simplicity let us put 
$\underline{\sigma}=\sigma_{i_1}...\sigma_{i_r}$, 
and for $x=\beta^2/N$ we have
\bes
\omega_{x=\beta^2/N}(\underline{\sigma})=
\frac{1}{Z_{N}(\beta)}
\sum_{{\sigma}}\exp(\frac{\beta}{\sqrt{N}}(\sum_{i<j}^{1,N}J_{ij}
\sigma_i\sigma_j +\sum_i J_i\sigma_i))\underline{\sigma}\ . 
\ees
Introducing a sum over $\sigma_{N+1}$ at the numerator and at the
denominator (which is the same as multiplying and dividing by $2^N$
because there is no dependence on $\sigma_{N+1}$), and making the
transformation  $\sigma_i\rightarrow \sigma_i\sigma_{N+1}$, the
variable $\sigma_{N+1}$ appears at the numerator and it is possible
to construct the state $\omega$ for $N+1$ particles:
\bes 
\omega_{N,x= \beta^2/N}(\sigma)=
\omega_{N+1}(\sigma\sigma_{N+1}^r) +O(\frac{1}{N})\ . \ \ \ \Box
\ees
\textbf{Proof of Theorem \ref{qqq}}
A fillable polynomial will be taken of the form
\bes
\label{101} Q_{ab}=\sum_{ij}\frac{\sigma_i^a\sigma_j^b}{N^2}
Q_{ij}(\sigma)\ ,
\ees
so that the $a,b$ are the non-filled replicas, fillable by $q_{ab}$,
and the filled replicas are included in the $Q_{ij}$'s.
We have
\bes
\label{102} \langle Q_{ab} \rangle_t =
\frac{1}{N^2}\mathbb{E}[\sum_{ij}\Omega_t(\sigma_i^a\sigma_j^b
Q_{ij}(\sigma))]=  \frac{1}{N^2}\mathbb{E}[\sum_{ij}
\omega_t(\sigma_i^a)\omega_t(\sigma_j^b)\Omega_t(Q_{ij})]\ . 
\ees
We will now take $x=\beta^2/N$ and use the previous lemma to get
%
\bes\label{103} 
\omega_{x=\beta^2/N}(\sigma_i^a)=\omega
(\sigma_i^a\sigma_{N+1}^a) + O(\frac{1}{N}) 
\ees
while the remaining part product state $\Omega_x$ continues 
to act on an even number of each replica and it is not modified:
\bes\label{104} 
\Omega_{x=\beta^2/N}(Q_{ij})=\Omega(Q_{ij})\ . 
\ees
Hence
\bes\label{105} 
\omega(\sigma_i^a\sigma_{N+1}^a)
\omega(\sigma_i^b\sigma_{N+1}^b)\Omega(Q_{ij})=
\Omega(\sigma_i^a\sigma_j^b \sigma_{N+1}^a\sigma_{N+1}^b Q_{ij})\ .
\ees
Using the dummy summation variable $\sigma_{N+1}$
\cite{Glocarno} we can also sum over all dumb indexes from 1 to $N$ 
and divide by $N$,
because the involved terms are of order
$O(\frac{1}{N})$ and become irrelevant in the $N \to \infty$
limit. Therefore
\bes\label{106} 
\langle Q_{ab}\rangle_{x=\beta^2/N}=
N^{-3}\mathbb{E}[ \sum_{ijk}\Omega(\sigma_k^a \sigma_k^b \sigma_i^a
Q_{ij}\sigma_j^b)]+O(\frac{1}{N}) 
\ees
and in the thermodynamic limit we complete the proof. $\Box$

If we now let $x\to 0^{+}$ as $\beta^{2}/N$, $N\to\infty$ we immediately get
\begin{corollary}\label{positive} Let $\beta >\beta_{c}$, then
$$
\lim_{N\to\infty}\langle q \rangle_{x=\beta^{2}/N} > 0\ .
$$
\end{corollary}
{\bf Proof}
This is the simple case of the fillable polynomial $Q=q$ in the previous
theorem, which offers
$\lim_{N\to\infty}\langle q \rangle_{x=\beta^{2}/N}= 
\lim_{N\to\infty}\langle q^{2} \rangle \geq 0$, 
which becomes a strict inequality below the critical point. $\Box$

The corollary shows that the the non-negativity of the average of the 
overlap can be easily proven with a simple cavity argument 
with no need of the Ghirlanda-Guerra relations, which are
instead needed to prove the non-negativity of the overlap
with probability one \cite{talabook}. 

We are interested in studying the behavior 
of the overlap at the critical point, where the overlap begins
to fluctuate. In order to do so, we will have to pay a price:
we need to provide a more involved proof of an even weaker statement 
than that of the previous corollary. This will pave the way to 
a new proof of the corollary which also offers a control of the critical
behavior of the overlap.

\subsection{The expansion and the spin-flip symmetry breaking}

We are interested in obtaining an expansion of the overlap $\langle
q_{12} \rangle$ in a neighborhood of the critical inverse
temperature $\beta_c$. 



The right way to perform the expansion is indicated by the next
\begin{proposition}\label{guerra}
Let $\phi_s$ be measurable with respect to the $\sigma$-algebra
$\mathcal{A}_s$ generated by the overlaps among $s$ replicas, then
the following streaming equation for $\phi_{s}$ holds: 
\bes
\partial_x \langle \phi_s \rangle_{x} =N \langle \phi_s (\sum_{a,b}q_{a,b}
-s\sum_{a=1}^sq_{a,s+1}+ \frac12 s(s+1)q_{s+1,s+2})\rangle_{x}
\ees
where all the averages are considered at any given size of the
system $N$.
\end{proposition}
This proposition is a straightforward generalization of the streaming equation
proved in \cite{gsum}. Throughout the section all the averages
are understood to be taken at fixed finite $N$.
If we use the proposition above in the simple
case $\phi=q$, we obtain immediately
\be\label{tre}
\partial_x \langle q_{12} \rangle_{x} = N
\langle q^2_{12} -4 q_{12}q_{23} +3 q_{12}q_{34} \rangle_{x}\ ,
\ee
which at $x=0$ reduces to
$N\langle q_{12}^{2} \rangle$ (with a violation
of Ghirlanda-Guerra identities \cite{gg} in the thermodynamic limit) 
and explodes as $N\to \infty$,
while it remains finite when $x > 0$.
We will get back to this later on.

It is from the above identity (\ref{tre}) that we will deduce a
weaker form of Corollary \ref{positive}, as stated in the next
\begin{theorem}\label{weaker}
 Let $x=y^{2}/N$ and $\beta > \beta_{c}$. Then, for sufficiently
small but non-vanishing $y$
$$
\lim_{N\to \infty}\langle q_{12}\rangle_{x=y^2/N}
> 0\ .
$$
\end{theorem}
{\bf Proof}
We are going to give place to an iterated expansion,
about the critical point $\beta_{c}$, by means
of a repeated use of Proposition \ref{guerra},
relying also on Proposition \ref{xxx}.
The right hand side
of (\ref{tre}) consists of three terms. The first term is (the
average of) the filled polynomial $\langle q_{12}^2 \rangle_x$,
which, as we noticed, does not depend on $x$ and hence
\be\label{integrale}
\langle q_{12} \rangle_{x} = N\left( \langle
q_{12}^2 \rangle x -4 \int_0^x dx'\langle q_{12}q_{23} \rangle_{x'}
+3 \int_0^x dx' \langle q_{12}q_{34} \rangle_{x'}\right)\ .
\ee
We want to show that the two terms under integral are of order higher
than the first in $x$. We will perform this task for the first term (the case of
the second is analogous) using again Proposition \ref{guerra}:
\begin{eqnarray}\label{g2}
\partial_x \langle q_{12}q_{23} \rangle_{x}& = & N
\langle q_{12}q_{23}q_{13}\rangle_x + O(x^3)\ .
\end{eqnarray}
Recalling the mentioned properties of filled polynomials, we immediately see
that equation (\ref{g2}) gives
\bes
\langle q_{12}q_{23}
\rangle_{x}=N \int_0^x \langle q_{12}q_{23}q_{13}\rangle_{x'} dx' =
N\langle q_{12}q_{23}q_{13}\rangle x\ .
\ees
so that consequently
$$
\int_0^x dx'\langle q_{12}q_{23} \rangle_{x'} = \int_0^x dx' N
\langle q_{12}q_{23}q_{13}\rangle_{x'}x' + O(x^3)\ .
$$
Therefore the leading contribution from $\langle q_{12}q_{23}
\rangle_{x}$ in (\ref{integrale}) is $\langle
q_{12}q_{23}q_{13}\rangle N^2x^2$, i.e. a contribution of order two
in $x$. 
An identical consideration tells us that the
contribution from the  last term in equation (\ref{integrale}) is of
three in $x$. 
Neglecting terms of order $x^4$ or
higher one thus gets
\begin{multline}\label{expansion}
 \langle q_{12} \rangle_{x}= N \langle q^2_{12}
\rangle x -2 N^2 \langle q_{12}q_{23}q_{13} \rangle x^2 - \\
-\frac{4}{3}N^3 \langle q^2_{12}q^2_{23} \rangle x^3 + N^3
 \langle q^2_{12}q^2_{34} \rangle x^3 +
 6 N^3 \langle q_{12}q_{23}q_{34}q_{14} \rangle x^3 +\cdots
\end{multline}
Notice that for a given strictly positive $x$, the above expansion
diverges below the critical temperature (we will comment further on
this at the end of the section) as $N\to\infty$, so that such an
expansion makes sense only if $x\to 0^{+}$ as $N\to\infty$ (this is
true within our spin-flip symmetry breaking approach, and we will see
later how to bypass this restriction studying the gauge symmetry
breaking). Let hence put $x=y^{2}/N$, as announced. Then we can
rewrite the previous expansion (\ref{expansion}) as
\begin{multline*}\label{expansion}
\langle q_{12} \rangle_{x}= \langle q^2_{12} \rangle y^2 -2
\langle
q_{12}q_{23}q_{13} \rangle  y^4 -\\
-\frac{4}{3} \langle q^2_{12}q^2_{23}  \rangle  y^6+ \langle
q^2_{12}q^2_{34} \rangle
  y^6 +6 \langle q_{12}q_{23}q_{34}q_{14} \rangle  y^6+\cdots\ .
\end{multline*}
For sufficiently small $y$ the sign of the left hand side is
determined by the first term in the expansion, which is zero above
the critical temperature but strictly positive below (for otherwise
the replica symmetric solution would hold at low temperatures, which
is not possible. See \cite{toninelli,guerraAT} for details). Hence
\bes 
\lim_{N\to \infty}\langle q_{12}\rangle_{x=y^2/N}>0
\ ,\ \beta > \beta_{c}\ ,
\ees 
while we already emphasized that \bes
\lim_{N\to \infty}\langle q_{12}\rangle_{x=0}\equiv 0 \ .\ \Box \ees
In other words, the limit depends on the way it is performed, and a
different way to identify the transition is stating that the two
limits $x\to 0^{+}$ and $N\to \infty$ (which we take
simultaneously) cannot
be interchanged below the critical temperature.
Looking at (\ref{tre}) - and its consequent
expansion (\ref{expansion}) - we can see that this transition we
just proved is connected with the violation of Ghirlanda-Guerra
identities \cite{gg}, when $\beta > 1$ and $x=0$ {\sl vs} $x > 0$
(we will deepen this further in a work still in progress at the moment).

\section{Critical indices}\label{exp}



We saw that, due to the gauge symmetry, the choice $x=\beta^{2}/N$
is equivalent to the addition of a new spin, say numbered $N+1$
(see for instance Lemma \ref{n+1}). 
So could we choose in the previous section $y=\beta$ 
(which makes the external field a proper cavity field), we would in some sense
identify the transition by getting two different results when
performing the limit as $N\to\infty$ 
of the average of the overlap for $N$ or $N+1$ spins. 
Unfortunately in the previous section $y$ has to be small enough
and this interpretation in general is not possible within the proof
we provided. Still, the expansion we illustrated can be adapted 
in such a way to recover the same result not limited to small $y$,
and to obtain the critical exponents too. 

Roughly speaking, the idea is that even if $y$ is not necessarily small,
the overlaps will be small anyways, as they are always bounded by one, and
they are identically zero above the critical temperature, at which they
start deviating from zero with continuity. This fact, together with 
an explicit use of the gauge symmetry (as opposed to mere 
spin-flip) will allow for the reformulation of the expansion we need.



\subsection{Lower bound for the first critical index}

Critical indices are needed to characterize singularities of the
theory at the critical point and are related to
the overlap correlation functions.

Let us introduce the expansion parameter 
$$\tau=(\beta^2-1)/2\ ,
$$ 
and let us focus on the averaged squared overlap 
$\langle q_{12}^2 \rangle$.
Assuming continuity in $\beta$ at the critical point, let us also
assume a behavior of the form 
\bes \langle q_{12}^2 \rangle \sim C
\tau^\gamma\ . 
\ees 
Once $\gamma$ is known, we can get all the other
critical exponents for the higher order correlation functions.

We will use convexity of the free energy density to obtain a bound
for $\gamma$ at a given $C$. We know that 
\bes 
\partial_{\beta}^2 \alpha_N(\beta) \geq 0
\ees 
which means that the internal energy density
$u_N(\beta)=-\partial_{\beta}\alpha_N(\beta)$ is a decreasing
function of $\beta$ 
\bes
\partial_{\beta} u_N(\beta)\leq 0\ .
\ees 
Therefore for every $\beta>1$, we have $u(\beta) \leq u(1)$.
From now on we will restrict ourselves to the case
$\beta>1$. We can thus write 
\bes
 -\frac{\beta}{2}(1-\langle q_{12}^2 \rangle) \leq
-\frac{1}{2}(1- \langle q_{12}^2 \rangle|_{\beta=1})=-\frac{1}{2}
\ees 
so that 
\be\label{epsilon} \langle q^2_{12} \rangle  \leq
\frac{\beta-1}{\beta}\ . 
\ee 
Using our assumption 
$\langle q_{12}^2\rangle = C\tau^\gamma +o(\tau^\gamma)$, 
the previous inequality (\ref{epsilon}) reads 
\be \label{bound} 
C\tau^{\gamma}
+o(\tau^\gamma) < \tau+o(\tau)\ . 
\ee 
Now,
\begin{itemize}
\item
if $\gamma <1$, condition (\ref{bound}) is not fulfilled. In fact, if
we divide by $\tau^\gamma$, the condition reads $C\leq
\tau^{1-\gamma}+o(1)$, and since $1-\gamma> 0$, in the limit
$\tau\to 0$ we would get $C\leq 0$, which is not possible;
\item
if $\gamma=1$, condition (\ref{bound}) becomes $C \leq 1+o(1)$,
which in the limit $\tau\to 0 $ gives $C\leq 1$.
\end{itemize}
Hence $\gamma\geq1$, and in particular if  $\gamma=1$ then $0<C<1$.
We are about to prove that $\gamma=2$.

\subsection{A revisited variational approach}

Let $p$ be a generic filled overlap polynomial. With an abuse of
notation, we follow \cite{barra} and consider the pressure,
expanded in overlap polynomials, as a function of the polynomials
with an explicit dependence on the inverse temperature: 
$\alpha(\beta)=\alpha(\langle p(\beta) \rangle,\beta)$. 
More precisely (see \cite{barra}) 
$$
\alpha(\beta, \langle p(\beta)\rangle )= \ln2 +
\frac{\beta^2}{4}[1+(1-\beta^2)\langle q_{12}^2 \rangle] +
\frac{\beta^6}{3}\langle q_{12}q_{23}q_{13} \rangle + O(\beta^8)
$$
and we are interested in computing its derivative
with respect to $\beta^2$. As
explained in details in \cite{barra}, the total derivative of the
$\alpha(\beta)$ with respect to $\beta^2$  is proportional to the
internal energy of the model, and as $\alpha(\beta)=
-\beta f(\beta)$, where $f(\beta)$ is the free energy density,
we derive with respect to $\beta^2$ rather than to $\beta$. This
is just matter of tastes, after all, because it would change 
just a factor in front of our expressions below.
We have
$$
\frac{d \alpha(\beta, \langle p(\beta) \rangle)}{d \beta^2} =
\frac{1}{2\beta}\frac{d \alpha(\beta,\langle p(\beta) \rangle)}{d
\beta}= \frac{1}{4}(1- \langle q_{12}^2 \rangle)
$$
and this expression needs to be equal to
$$
\frac{d \alpha(\beta, \langle p(\beta) \rangle)}{d \beta^2} =
\partial_{\beta^2}\alpha(\beta,\langle p(\beta) \rangle) +
\sum_{p}\frac{\partial \alpha (\beta, \langle p(\beta)
\rangle)}{\partial \langle p(\beta) \rangle}\frac{\partial \langle
p(\beta) \rangle}{\partial \beta^2}\ .
$$
By means of some tedious calculations
that we will not report (we rather refer the interested reader
to \cite{barra}), we get
$$
\partial_{\beta^2}\alpha(\beta,\langle p(\beta) \rangle)
= \frac{1}{4}(1- \langle q_{12}^2 \rangle)\ ,
$$
which imposes the following constraint
$$
\sum_{p}\frac{\partial \alpha(\langle p \rangle,\beta)}{\partial
\langle p \rangle}  \frac{\partial \langle p \rangle}{\partial
\beta^2}=0\ ,
$$
where the sum on $p$ denotes the sum over all the overlap
correlation functions appearing in the expansion.
From this equation, one can deduce \cite{barra}
the Aizenman-Contucci polynomials (AC) through an approach 
different from stochastic stability (see \cite{ac}). Here we want to
employ such an approach to 
derive all the critical exponents of the theory
within an iterative scheme.

Let us take the first two non trivial orders (with $\langle
q_{12}^2 \rangle$, $\langle q_{12}q_{23}q_{13} \rangle$) 
and write,
neglecting terms of higher order than $O(q^4)$
\be\label{principiomio} 
\frac{\partial \alpha(\beta)}{\partial
\langle q_{12}^2 \rangle} \frac{\partial \langle q_{12}^2
\rangle}{\partial \beta^2} + \frac{\partial \alpha(\beta)}{\partial
\langle q_{12}q_{23}q_{13} \rangle}\frac{\partial \langle
q_{12}q_{23}q_{13} \rangle}{\partial \beta^2}=0\ . 
\ee 
From
(\ref{integrale}) and the first AC relation 
$$\langle
q_{12}^4 \rangle -4 \langle q_{12}^2 q_{23}^2 \rangle +3\langle
q_{12}^2 q_{34}^2\rangle=0
$$ 
we get, when $x=\beta^{2}/N$,
\begin{eqnarray*}
\langle q_{12} \rangle &=& 
\langle q_{12}^2
\rangle \beta^2 -2 \langle q_{12}q_{23}q_{13} \rangle \beta^4 
-\frac{1}{3}\langle q_{12}^4 \rangle + 6 \langle
q_{12}q_{23}q_{34}q_{14} \rangle\beta^6 + O(q^5)\\
\langle q_{12}q_{23} \rangle &=&
\langle q_{12}q_{23}q_{13} \rangle \beta^2 + \langle
q_{12}^2q_{23}^2 \rangle \beta^4  -3 \langle
q_{12}q_{23}q_{34}q_{14} \rangle\beta^6+ O(q^5)\ .
\end{eqnarray*}
Recalling Proposition \ref{qqq} we also obtain
\begin{eqnarray} 
(\beta^2-1) \langle
q_{12}^2 \rangle &\!\!\!=\!\!\!& 
2\beta^4 \langle q_{12}q_{23}q_{13}\rangle +
\frac{\beta^6}{3} \langle q_{12}^4 \rangle - 6\beta^6 \langle
q_{12}q_{23}q_{34}q_{14} \rangle \label{beta-1-uno} \\
(\beta^2-1) \langle q_{12}q_{23}q_{13} \rangle &\!\!\!=\!\!\!& 
-\beta^4
\langle q^2_{12}q^2_{23} \rangle +
3 \beta^4 \langle q_{12}q_{23}q_{34}q_{14}\rangle\ .  \label{beta-1-due}
\end{eqnarray}
At this point the AC relations are not enough, and we need
their non-linear extension expressed in the 
following Ghirlanda-Guerra identities \cite{gg} 
\bes 
\langle q_{12}^2q_{13}^2 \rangle = \frac{1}{2}
\langle q_{12}^4 \rangle + \frac{1}{2} \langle q_{12}^2 \rangle ^2\
,\ \langle q_{12}^2q_{34}^2 \rangle = \frac{1}{3} \langle q_{12}^4
\rangle + \frac{2}{3} \langle q_{12}^2 \rangle ^2\ , 
\ees 
to deduce from (\ref{beta-1-uno})-(\ref{beta-1-due})
the expression of the overlap correlation functions of higher orders 
in terms of the two-replica overlap 
\begin{eqnarray*} 
\beta^2 \langle
q_{12}q_{23}q_{13}\rangle &=& \tau \langle q_{12}^2 \rangle +
\frac{1}{3} \langle q_{12}^4\rangle +\frac{1}{2} \langle q_{12}^2
\rangle^2+O(q^5)\ , \\
\langle q_{12}q_{23}q_{34}q_{14}\rangle &=& 
\frac{1}{6} \langle
q_{12}^4\rangle + \frac{1}{6} \langle q_{12}^2\rangle^2 +
\frac{2}{3}\tau^2 \langle q_{12}^2 \rangle+O(q^5)\ . 
\end{eqnarray*}
Now, calling $\langle q_{12}^2 \rangle = m^2(\beta)$ and noting that
$\langle q_{12}q_{23}q_{13} \rangle \sim \tau m^2(\beta) \sim C
\tau^{\gamma+1}$, we substitute $m^2(\beta)$ into 
(\ref{principiomio}) and get 
\bes 
-\frac{\tau}{2}\frac{d
(m^2)}{d\beta^2} + \frac{1}{3} \frac{d (\tau m^2)}{d \beta^2} =0
\ees 
where the factors $\frac{1}{2}$ and
$\frac{1}{3}$ come from the form of the expansion of the free energy
(see \cite{barra} for details).
The derivative of second term in the left hand side above gives
$$
\frac{d(\tau m^2)}{d \beta^2}= \frac{1}{2}m^2 + \tau
\frac{d(m^2)}{d\beta^2}\ ,
$$
so that we obtain a differential equation trivially solvable by
separation of variables: \bes \frac{d \beta^2}{\tau(\beta)}= \frac{d
m^2}{m^2}= d \log(m^2(\beta))\ . \ees Integrating against
$d(\beta')^2$ yields
$$
\int_{\beta_1^2}^{\beta^2} \frac{2 d (\beta')^2}{\beta^2 -1} =
\int_{\beta_1^2}^{\beta^2} d \log m^2(\beta) \rightarrow
\log(\beta^2-1)^2 = \log(m^2(\beta))+ C
$$
where $\beta^2>\beta^2_1>1$ and $C$ will depend on $\beta_{1}$. So
we finally have \bes \langle q^2_{12} \rangle = m^2(\beta)=
C(\beta^2-1)^2 + O((\beta^2-1)^3) \sim C \tau^2\ . \ees Therefore
the critical index for the first overlap correlation function is
$\gamma = 2$. From this, we immediately deduce
\begin{eqnarray*}
\langle q_{12}q_{23}q_{13} \rangle &\sim& C \tau^3 \\
\langle q_{12}q_{23}q_{34}q_{14} \rangle &\sim& C' \tau^4 \\
\langle q_{12}^4 \rangle &\sim& C'' \tau^4
\end{eqnarray*}
and so on.

Notice that the steps we illustrated in this section 
provide an improvement of Theorem \ref{weaker}, as
they allow to extend its statement up to $y=\beta$,
thus recovering Corollary \ref{positive}.

\section{Conclusions and Outlook}\label{final}

In this paper we have shown that the SK model exhibits the phase
transition predicted by Parisi in terms of the discontinuous
response of the overlap to the presence of and external field. A
simple iterated power expansion of the overlaps within the cavity
method allowed for the proof of the results. The techniques we used
are extensions of those introduced and employed in \cite{barra} and
\cite{bds1}. As a result the control near the critical point of the
overlap discontinuity in the $(\beta,x)$ plane follows, together
with the computation of the critical exponents of the overlap
correlation functions. 

\section{Acknowledgment}
The authors are pleased to thank Francesco Guerra for his constant
priceless scientific support. We are also very grateful to Peter Sollich,
Silvio Franz and Andrea Pagnani for fruitful discussions. The authors
thank an anonymous referee for useful suggestions.


\end{document}